\documentclass[twocolumn,prl,aps,showpacs]{revtex4}

\usepackage{graphicx}
\usepackage{dcolumn}
\usepackage{amsmath}
\usepackage[french]{babel}
\usepackage[latin1]{inputenc}

\begin{document}

\title{On the physical meaning of the gauge conditions
 of Classical Electromagnetism :\\ the hydrodynamics analogue viewpoint}

\author{Germain Rousseaux}
\affiliation{Physique et M\'{e}canique des Milieux
H\'{e}t\'{e}rog\`{e}nes,\\ UMR 7636 CNRS-ESPCI,\\ 10,
Rue Vauquelin, 75231 Paris Cedex 05, France\\
(germain@pmmh.espci.fr)}
\date{\today}

%\doublespacing

\begin{abstract}

Based on an analogy between Fluid Mechanics and Electromagnetism,
we claim that the gauge conditions of Classical Electromagnetism
are not equivalent contrary to the common belief. These "gauges"
are usually considered as mathematical conditions that one must
specify  in order to solve any electromagnetic problem. Here, the
author shows that these conditions are physical constraints which
can be interpreted as electromagnetic continuity equations. As a
consequence, light cannot be considered as a pure transverse wave
in vacuum from the point of view of the potentials. We discuss the
(lack of) meaning of gauge transformations.

\end{abstract}


\maketitle

\section{Introduction}

In Classical Electromagnetism, the generalized momentum {\bf p} of
a particle with mass m and charge q moving at a velocity {\bf v}
in a vector potential {\bf A} is \cite{Jackson} : ${\bf p} = m
{\bf v} + q {\bf A}$. Hence, the vector potential can be seen as
the electromagnetic impulsion (per unit of charge) of the field.
For example, induction phenomena are due to the transfer of
momentum from the field to the charge via the vector potential.
Simply speaking, the vector potential is a kind of velocity up to
a factor q/m. There is a long history of papers
(\cite{Konopinski,Taylor} and references therein) and of books
which advocate forcefully a physical interpretation to the vector
potential (\cite{Maxwell,Feynman,Tonomura,Mead} and references
therein). One can define the vector potential at a point M as the
mechanical impulsion that an external operator must furnish to a
unit charge in order to bring it from infinity (where the vector
potential vanishes far from the currents) to the point M. The
generalized energy $\epsilon$ of the same particle in a scalar
potential V is \cite{Jackson} : $\epsilon=mv^2 /2 + qV$. Hence,
the scalar potential can be seen as the potential energy (per unit
of charge) of the field. For example, an electron is accelerated
in an electron gun and its gained energy per unit charge is the
scalar potential. Of course, the potentials are defined up to a
constant and the experimentalist sets by convention the scalar
potential of a plate in the electron gun to zero for instance.
Similarly, one can define the scalar potential at a point M as the
mechanical energy that an external operator must furnish to a unit
charge in order to bring it from infinity (where the scalar
potential vanishes far from the charges) to the point M.

Usually, in order to solve a problem in Electromagnetism, one must
specify what is called a gauge, that is a supplementary condition
which is injected in the Maxwell equations expressed in function
of the electromagnetic potentials. Two gauge conditions were
introduced in Classical Electromagnetism \cite{Jackson} : $
\nabla\cdot{\bf{A}}=0$ which is the Coulomb gauge, used for
example in magnetostatics, and : $ \nabla\cdot{\bf{A}} + 1/c_L^2
\partial_t V=0$ which is the Lorenz gauge.
Here, $c_L$ is the velocity of light \cite{Jackson} : $c_L=\sqrt
{1 /\mu_0 \epsilon_0}$ where $\mu_0$ and $\epsilon_0$ are
respectively the permeability and the permittivity of the vacuum.
It is common to say that these gauge conditions are mathematical
conveniences that lead to the same determination of the
electromagnetic field. In this context, the choice of a specific
gauge is motivated  from its conveniences in calculations !

We would like to underline that these gauges may not be equivalent
:
\begin{itemize}
\item From the mathematical point of view, the Coulomb gauge is
the approximation of the Lorenz gauge in the stationary case
-which is a well known result - but also when the velocity of
light is taken to be infinite (what this paper will
demonstrate...).

\item From the physical point of view, the gauges can be seen as
electromagnetic continuity equations. To understand this last
point, one can use the following analogy with hydrodynamics.
\end{itemize}

\section{The analogue proof of the non-equivalence between the gauge conditions}

In order to solve a problem in fluid mechanics, one must specify a
physical constraint which tells us if the fluid flow is
compressible or not. The incompressibility constraint reads
\cite{Guyon} : $ \nabla \cdot{\bf{u}}=0$ whereas the
compressibility constraint is \cite{Guyon} : $ \nabla \cdot{\bf
u}+ 1/\rho  D_t \rho=0$ where $ D_t=\partial _t()+({\bf
u}\cdot\nabla)() $ is the so-called total derivative, $\bf{u}$ is
the velocity of a fluid particle and $\rho$ its density. If the
flow is not stationary and if one considers acoustic waves which
are perturbations of the pressure, the density and the velocity of
the fluid around a basic state (subscript 0) $p=p_0+\delta p$,
$\rho=\rho_0+\delta \rho$ and $\bf{u}=\bf{0}+\delta{\bf u}$, one
can evaluate the velocity of sound by the following formula
\cite{Guyon} : $ c_S=\sqrt {\partial p /\partial \rho}=\sqrt {1/
\rho \kappa}$ where $\kappa$ is the compressibility of the fluid.
The compressibility constraint becomes : $ \nabla \cdot\delta{\bf
u}+1/c_S^2 \partial _t(\delta p / \rho_0)=0$ which has a form
equivalent to the Lorenz gauge. If the velocity of sound tends to
infinity, one recovers the incompressibility constraint. The new
result is that the Coulomb gauge would imply that the velocity of
light tends to infinity in a time dependent problem when
propagation is absent as in hydrodynamics \cite{Rousseaux}.
Moreover, if the flow is stationary, the compressibility
constraint reduces to the incompressibility constraint which is
analogous to the Coulomb gauge.

\section{The experimental proof of the non-equivalence between the gauge conditions}

Now, one can read in every textbooks of electromagnetism that we
can describe propagation of potential waves in either the Coulomb
or the Lorenz gauge because in any case the propagation of the
electromagnetic waves remains unchanged... We will show that the
Coulomb constraint cannot describe propagation at finite speed but
instantaneous propagation in a coaxial cable.

What does it mean experimentally that a quantity propagates
instantaneously? Imagine the following experiment. Let's take a
coaxial cable. One can relate it to a function generator which
delivers for instance a scalar potential pulse of whatever shape :
square, triangular... Experimentally, the scalar potential seems
to propagate instantaneously in a short coaxial cable of one meter
long. We are in the so-called quasi-static limit where quantities
are time-dependent but do not propagate (with the help of the
analogy, one understand that we should use the Coulomb
constraint). Experimentally, the scalar potential does not
propagate instantaneously in a long coaxial cable of one hundred
meter long because we are able to detect with an oscilloscope a
time delay introduced by the finite speed propagation of a pulse
of scalar potential between the entry and the exit of the cable.
This last experimental fact is in contradiction with the assertion
that we can use Coulomb gauge to describe propagation because the
scalar potential is solution of a Laplace equation in this gauge
that is, it must propagate instantaneously \cite{Jackson}. From
the analogy, one concludes that we should use the Lorenz
constraint to describe propagation and not the Coulomb constraint.
Of course, one also uses Coulomb constraint in the
time-independent case. A close look to the range of validity of
the so-called quasi-stationary approximation ($c_L$ is infinite)
permits to understand that there is no contradiction with the
above statement concerning the fact that the potentials and the
fields can or can not propagate depending on the problem...

To conclude this part, whatever the potentials are undetermined or
not, the so-called gauges conditions seem to be physical
constraints which would tell us if the velocity of light is a
relevant parameter or not (that is finite or not) and not
mathematical conditions to fix the potentials. Indeed, depending
on what type of phenomena you are studying, some imply that the
velocity of light is finite and some other do not. More precisely,
is there a consistent galilean electromagnetism ($c_L$ is
infinite) coexisting with a relativistic electromagnetism ($c_L$
is finite) ? This question was addressed and answered for the
fields by L\'{e}vy-Leblond \& Le Bellac \cite{Levy} and it was
revisited recently by Holland \& Brown \cite{Holland}. Our paper
extends these last works for the gauge conditions.

\section{The mathematical proof of the non-equivalence between the gauge conditions}

As a matter of fact, L\'{e}vy-Leblond \& Le Bellac have shown that
the full set of Maxwell equations has two well defined galilean
limits which they called the magnetic limit (used for example in
Ohmic conductors and in Magnetohydrodynamics
(\cite{Moreau90,Melcher}) : also called the magneto-quasistatic
approximation) and the electric limit (used for example in
dielectrics and in Electrohydrodynamics
(\cite{Moreau73,Melcher,Castellanos}): also called the
electro-quasistatic approximation). The two limits are obtained by
taking the velocity of light as infinite. Contrary to mechanics
which allows only one galilean limit, the two limits of
electromagnetism come from the fact that $c_L=\sqrt {1 /\mu_0
\epsilon_0}$ can tend to infinity if either $\mu_0$ or
$\epsilon_0$ tends to zero separately. For example, the magnetic
limit is the result of keeping $\mu_0$ constant during the process
while $\epsilon_0$ tends to zero.

Moreover, L\'{e}vy-Leblond \& Le Bellac have derived the galilean
transformations for the potentials \cite{Levy}. In the magnetic
limit (${\bf u}<<c_L$ and $V<<c_L.|{\bf A}|$), they read : ${\bf
A^*}={\bf A}$ and $ V^*=V-{\bf u}.{\bf A} $ whereas in the
electric limit (${\bf u}<<c_L$ and $V>>c_L.|{\bf A}|$): ${\bf
A^*}={\bf A}-{\bf u}/c_L^2 V$ and $V^*=V $. Now, if we apply the
limiting process used by these authors (${\bf u}<<c_L$ and
$V<<c_L.|{\bf A}|$ or $V>>c_L.|{\bf A}|$) to the Lorenz gauge
which, we know, is Lorentz invariant ($c_L$ is finite), we find
that the Lorenz gauge resumes to the Coulomb gauge in the magnetic
limit and that the Lorenz gauge remains the same in the electric
limit. The Lorenz (Coulomb) gauge is now covariant with respect to
the "electric" ("magnetic") transformations of the potentials. The
Coulomb gauge is the only possible constraint that we can apply
when we deal with Ohmic conductors or in Magnetohydrodynamics that
is within the range of the magnetic limit. The Coulomb gauge
cannot apply in the electric limit as well as in relativistic
electromagnetism which was not recognized before. The important
point is that the Coulomb gauge is obtained mathematically by a
limiting process from the Lorenz gauge and is not independent of
the Lorenz gauge. We clearly state that it is hence forbidden to
plunge the Coulomb gauge which is galilean into the full set of
Maxwell equations which are relativistic contrary to what is
stated in almost all the textbooks. The Lorenz gauge describes
both relativistic electromagnetism and galilean electromagnetism
within the electric limit and it cannot apply in the magnetic
limit.

Once again, the analogy can help us to grasp the underlying
physics. If a flow is said to be incompressible, the velocity of
sound is considered to be infinite. More precisely, the
compressibility of the fluid tends to zero while the density is
kept constant. Moreover, we characterized usually media where
waves propagate by using the concept of impedance which for an
acoustic wave is $Z_s=\rho_0c_s$ and for a light wave is $Z_L=\mu
_0c_L$. Hence, $\mu_0$ is the analogue of $\rho_0$. Now, we can
remark easily that the magnetic limit is the analogue of an
incompressible flow while there is no mechanical counterpart for
the electric limit. One understands why the Coulomb gauge is the
only gauge which does apply in Ohmic conductors within the
magnetic limit which are analogous to Newtonian fluids in
incompressible flow \cite{Rousseaux}. Recently, Brown \& Holland
\cite{Brown} have shown that the Schroedinger equation which is a
galilean equation was only coherent with the use of the magnetic
limit which explains why we use the Coulomb gauge with this
equation when dealing with an electron in a vector potential.

One century ago, H.A. Lorentz noticed that the electromagnetic
field remains invariant ($\bf{E'}=\bf{E}$ and $\bf{B'}=\bf{B}$)
under the so-called gauge transformations \cite{Okun} : ${\bf A'}=
{\bf A} + \nabla f$ and $V'=V-\partial _t f$ where $f(x,t)$ is the
gauge function. Hence, this indetermination is believed to be an
essential symmetry of Classical Electromagnetism \cite{Okun}. We
showed that the Coulomb and Lorenz gauges were not equivalent
because they must be interpreted as physical constraints that is
continuity equations. So, to make a gauge choice is not related to
the fact of fixing a special couple of potentials. Gauge
conditions are completely uncorrelated to the supposed
indetermination of the potentials. The gauge choice must be taken
with respect to the type of electromagnetism we study that is
relativistic or not by taking care also of the type of galilean
limit. What is the meaning of gauge transformations ? We believed
that it is only a structural feature (that is linearity) of the
definitions of the potentials from the fields. The potentials of
Classical Electromagnetism do have a physical meaning as recalled
in the introduction. If we defined the fields from the potentials
and not the contrary, the gauge transformations loose their sense.

As a conclusion, we propose to reject gauge transformations. Gauge
invariance is preserved but in a weaker sense : the potentials are
defined up to a constant. The proposed rejection of gauge
transformations is not new in the literature : it was foreseen by
L. de Broglie in the application of the principle of inertia of
energy in relativity \cite{Broglie}. More recently, A. Van Oosten
proposed a non-gauge-invariant theory of electromagnetism based on
the Fermi Lagrangian which is a valid alternative to the standard
approach as it makes the same experimental predictions
\cite{Oosten}.

\section{The Nature of Light}

Now, we can have a closer look to the way the propagation of light
is described usually. One can find for example in The Classical
Theory of Fields by Landau \& Lifshitz \cite{Landau} the following
description. Thanks to gauge invariance, one can take the Coulomb
gauge $\nabla\cdot{\bf{A}}=0$ and the assumption that the scalar
potential is zero in order to describe light propagation. As a
matter of fact, one obtains : ${\bf E}=-\partial_t {\bf A}-\nabla
V=-\partial_t {\bf A}$ and ${\bf B}=\nabla \times {\bf A}$. The
vector potential (so the fields) is solution of a propagation
equation. From our point of view, this derivation is misleading
because the Coulomb gauge is not Lorentz invariant and we
advocated in this paper that it can not describe propagation
($c_L$ is infinite). Indeed, if we apply a Helmholtz decomposition
to the Lorenz gauge ($c_L$ is finite), one finds :
$\nabla\cdot{\bf{A}}_{longitudinal}+ 1 /c_L^2
\partial_t V=0$ and independently :
$\nabla\cdot{\bf{A}}_{transverse}=0$. The magnetic field is : $
{\bf B}=\nabla \times {\bf A}_{transverse}$ with : $ \nabla \times
{\bf A}_{longitudinal}=0 $. The electric field writes : $ {\bf
E}=-\partial_t {\bf A}-\nabla V=-\partial_t {\bf A}_{transverse}$
with : $ \partial_t {\bf A}_{longitudinal}+\nabla V=0 $.

Indeed, concerning the nature of light, one can wonder if light
should still be considered as a transverse wave. As a matter of
fact, the potentials do have a physical meaning in Classical
Electromagnetism as recalled in the introduction. Moreover, we
gave a physical interpretation of the gauge and particularly of
the Lorenz gauge which implies by Fourier transform : $V=c_L A_x$
where x is the direction of propagation of a plane light wave in
vacuum ($k=\omega /c_L$). Hence, if we can say - as usual - that
the longitudinal electric and magnetic fields cancel ($E_x=-
{\partial_t(A_x cos(kx-\omega t))}- {\partial_x(V cos(kx-\omega
t))}=0$ and $B_x=(\nabla \times {\bf A})_x=0$ because $V=c_L
A_x$), the last equation shows that, under the Lorenz constraint,
the vector potential has a non-zero longitudinal component which
is a gradient. As pointed out by B. Leaf, the time-like and
longitudinal potential components constitute a Lorentz-covariant
null vector which is not amenable to boson quantization as the
transverse components \cite{Leaf}. So from the point of view of
the potentials from which the fields derive, light is neither a
transverse or a longitudinal wave : it is a composite wave...

Once again, one can understand the longitudinal propagation for
light with the sound analogy. By Fourier transformation of the
continuity equation for the fluid, one obtains : $\delta p/ \rho_0
=c_S \delta{u_x}$. One recalls that the propagation of sound waves
is vorticity-free ($B_x=0$) and that one gets the propagation
equations by combination of the continuity equation with the
linearized Navier-Stokes equation ($E_x=0$) \cite{Guyon} :
$\partial_t{\delta \bf u}+ \nabla{(\delta p / \rho_0)}=0$. The
longitudinal propagation for light is not in contradiction with
polarization experiments which do show that light can not be a
pure longitudinal wave but that the electric field is transverse
despite the fact that the vector potential has a longitudinal
component...

The longitudinal propagation of the potentials is also a feature
of electromagnetic waves in a coaxial cable with the difference
that the longitudinal vector potential is not a gradient in this
case \cite{Mead,Melcher}. The unconvinced reader could argue that
all the results regarding light can be derived without any
reference to the potentials. Formerly, it is right but there is an
implicit statement when we use the full set of Maxwell equations
to derive light propagation that is we consider the velocity of
light as finite. That's why we advocated in this paper that it is
equivalent to use the Lorenz gauge.

\section{Conclusions}

In conclusion, one can understand that the gauges express
electromagnetic continuity from a physical point of view based on
an analogy with hydrodynamics. From this analogy, we concluded
that the Lorenz gauge is more fundamental, in general, than the
Coulomb gauge which is an approximation for the stationary case
and for the time-dependent case when one neglects the propagation
of electromagnetic waves and more generally relativistic phenomena
within the magnetic limit. From the pedagogical point of view, the
analogy facilitates the use and understanding of the vectorial
operators and allows to find solutions of electromagnetic problem
much more readily in terms of hydrodynamics equivalent
\cite{Rousseaux}.

The author is fully aware that the conclusions of this paper are
controversial as they defy old-established opinions about the
non-physical character of the potentials as well as the so-called
gauge conditions. Anyway, it is the author's belief that
Electromagnetism cannot continue to be transmit to young
generations without understanding the fundamentals of this
discipline and in particular of the potentials which are the
primary quantity in relativity and quantum field theory. Let us
remind James Clerk Maxwell's own words :  [the vector potential]
is the mathematical quantity which can be considered as the
fundamental quantity of the electromagnetic theory
(\cite{Maxwell}, Vol.2, p. 187). It is funny to notice that
Maxwell used also the following expressions : electrotonic state,
electrokinetic momentum or electromagnetic momentum to designate
the vector potential...

\end{document}